\documentclass[aps,a4paper,preprint]{revtex4}%
\usepackage{amsfonts}
\usepackage{amsmath}
\usepackage{amssymb}
\usepackage{graphicx}
\usepackage{setspace}%
\providecommand{\U}[1]{\protect\rule{.1in}{.1in}}
\begin{document}
\title{Liquid-liquid capillary replacement in a horizontal geometry: universal
dynamics and replacement time}
\author{Julie Andr\'{e}$^{1,2}$ and Ko Okumura$^{1}$}
\affiliation{$^{1}$Physics Department and Soft Matter Center, Ochanomizu University, Japan}
\affiliation{$^{2}$Physics Department, \'{E}cole Polytechnique, France}
\date{\today}

\begin{abstract}
Capillary invasion of a liquid into an empty tube, which is called capillary
rise when the tube axis is in the vertical direction, is one of the
fundamental phenomena representing capillary effects. Usually, the tube is
actually filled with another pre-existing fluid, air, whose viscosity and
inertia can be practically neglected. In this study, we considered the effect
of the pre-existing fluid, when its viscosity is non-negligible, in a
horizontally geometry. This geometry is free from gravity and thus simpler
than the geometry of capillary rise. We observed the dynamics when a capillary
tube that is submerged horizontally in a liquid gets in contact with a second
liquid. An appropriate combination of liquids allowed us to observe that the
second liquid replaces the first without any prewetting process, thanks to a
careful cleaning of capillary tubes. Furthermore, we experimentally observed
three distinct viscous dynamics: (i) the conventional slowing-down dynamics,
(ii) an unusual accelerating dynamics, and (iii) another unusual dynamics,
which is linear in time. We developed a theory in viscous regimes, which
accounts well for the observations through a unified expression describing the
three distinct dynamics. We also demonstrated a thorough experimental
confirmation on the initial velocity of the replacement. We further focused on
the replacement time, the time required for the invading fluid to replaces
completely the pre-existing fluid in the horizontal geometry, which is again
well explained by the theory.

\end{abstract}
\maketitle

\section{Introduction}

Capillary rise, a key experiment in the field of interfaces, concerns liquids
brought into contact with a thin tube. Then liquids generally rises in the
tubes, up to a height resulting from a balance between capillarity and
gravity. The dynamics towards equilibrium is known to be divided into three
successive steps \cite{de2013capillarity}: an initial regime dominated by
inertia, an ensuing viscous regime, and a final gravity-viscous regime. The
corresponding dynamics are successively linear in time $t$ (meaning the rising
height $x$ scales with $t$), proportional to $t^{1/2}$ (the so-called
Lucas-Washburn law), and exponentially relaxing to the static height. In all
these regimes, air pre-existing in the tube can be ignored, but there are
cases where the fluid initially present in the tube has a non-negligible role.
For example, a recent study by Hultmark \cite{hultmark2011influence}, using
very long and thin tubes filled with air, evidences that the dynamics in the
viscous regime significantly deviates from the usual Washburn law. These
effects are amplified when the tube is initially filled by another liquid, as
it happens in oil recovery where pores full of oil are invaded by an aqueous
solution. Having in mind this field of applications, studies considered the
case of complex fluids \cite{elliott1967dynamic,hansen1971dynamic} and
tortuous geometries \cite{barenblatt1990Rocks,gerritsen2005modeling} -- all
situations where flows are forced instead of being spontaneous.

Spontaneous replacement of a liquid by another liquid was first addressed in
1946 by Eley \cite{eley1946dynamical}, who discussed this question both in
powders and in tubes. For a certain ratio of viscosities and pre-wetted tubes,
dynamics was observed to be accelerating, an unusual fact in capillary
phenomena, but this aspect was not studied further, the study focusing on
extracting the surface tension between the two liquids. Later on, Mumley
studied liquid-liquid rise, considering different prewetting states, and a
broad range of viscosities \cite{mumley1986kinetics}. A complete model was
derived (including both liquid viscosity and gravity) giving an implicit
solution, which was yet not compared with experiments. This liquid-liquid
capillary rise was revisited in 2016 by Walls \cite{walls2016capillary}, who
derived similar implicit dynamics and focused on the early dynamics under
gravity. Experiments confirmed the existence of an early linear regime, which
does not originate from an inertial dynamics but from the viscosity of the
displaced liquid.

Here we consider capillary invasion in a horizontal geometry, which we call
\textquotedblleft horizontal capillary replacement\textquotedblright\ to
emphasize the role of a pre-existing fluid to be replaced by another. Although
a horizontal geometry is simpler to model than capillary rise, there are very
few data are available on this subject. A theoretical and numerical study
\cite{budaraju2016capillary} recently investigated the liquid-liquid
horizontal capillary replacement, focusing on the influence of tapered tubes.
Dealing quickly with the case of a straight tube, the study predicted an
accelerated dynamics (observed in \cite{eley1946dynamical}) and a
viscous-linear linear dynamics, but no comparison with experiments was
presented. In the case where the displaced fluid is much less viscous than the
invading fluid, experimental data with liquid-air can be found in the paper by
Hultmark \cite{hultmark2011influence}. As for the inverse viscosity ratio in
horizontal geometry, there is surprisingly a complete lack of experimental
data in the literature.

We investigate liquid-liquid horizontal capillary replacement, for a broad
range of viscosity ratios. Thanks to the simple geometry, we analytically
solve the dynamics in a closed form, which unifies three distinct viscous
dynamics, which we all confirm experimentally: in addition to the conventional
viscous slowing-down dynamics, the theory captures the viscous-linear dynamics
and a viscous accelerating dynamics, which were mentioned theoretically in
\cite{walls2016capillary,budaraju2016capillary}. In this study, we
systematically compared the theory with experimental observations of the
accelerating capillary replacement.

We also demonstrate a thorough comparison between the theory and our
experimental data on the initial finite velocity, strengthening the previous
work from \cite{walls2016capillary} performed in vertical geometry. In
addition, the horizontal geometry provides a new variable of great interest:
the replacement time, that is, the time required to fully replace the liquid
initially present in the tube. This characteristic time can be relevant for
some industrial processes yet has not been discussed in the literature. In the
present study, we predict and verify experimentally the value of this
replacement time. We finally show how the initial presence of a small slug of
the invading liquid (a case of practical relevance) modifies the dynamics of replacement.

\section{Experiment}

\begin{figure}[h]
\includegraphics[width=\textwidth]{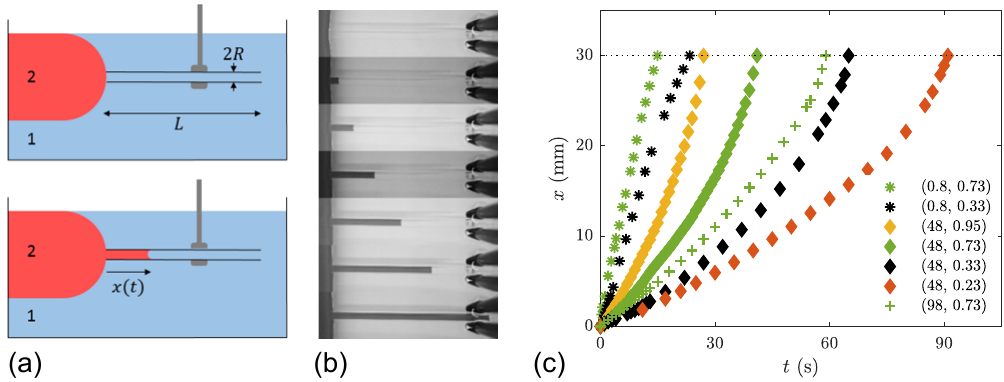}\caption{a) Schematic view of the
set up, with $R$ and $L$ the radius and length of capillary tube, liquid 1 is
silicon oil, liquid 2 is ethanol or water. $x(t)$ is the invasion length. b)
Successive snapshots of a typical experiment, with ethanol and silicon oil
with viscosity 48 mPa$\cdot$s, in a glass tube with $R=0.73$ and $L=30$ mm.
The first image is taken just before contact of the tube with the ethanol
drop. The interval between two images is 7 seconds. c) Invasion length $x$ as
a function of time $t$. The liquids are an ethanol solution ($\eta_{2}=1.4$
mPa$\cdot$s) and silicon oils of different viscosity $\eta_{1}$. The values of
$\eta_{1}$ (in mPa$\cdot$s) are $0.8$ (star marker), $48$ (diamond marker),
and $98$ (plus marker). The radii of the tube $R$ (in mm) are $0.23$ (orange),
$0.33$ (black), $0.75$ (green), and $0.95$ (yellow).}%
\label{Fig1}%
\end{figure}

We fill a shallow transparent container (radius 14.5 cm, depth 9 mm) with
liquid 1 and add a small amount of liquid 2 at the edge of the container, as
illustrated in Fig. \ref{Fig1} (a). Liquid 2 forms a lens of a millimeter
thickness, which we contact with a capillary tube immersed in liquid 1. The
diameter $2R$ ($R=0.10$ to 0.95 mm) and length $L$ ($L=12$ to 80 mm) of the
tube are respectively smaller than the thickness of the lens and the diameter
of the container. Tubes are made of glass and purchased from Hirschmann
Laborgerate GmbH \& Co. KG. Liquid 1 is a silicone oil with viscosity
$\eta_{1}$ (KF-96, Shin-Etsu Chemical Co., Ltd), which can be varied between
$0.8$ and 970 mPa$\cdot$s. Liquid 2 is an 80 \% ethanol aqueous solution
(sterilization ethanol IP, KENEI Pharmaceutical Co., Ltd), with viscosity
$\eta_{2}=1.4$ mPa$\cdot$s.

A special care was needed to clean capillary tubes to perform reproducible
experiments. The best performance was obtained by the following protocol: (1)
soak tubes in an aqueous solution of Hellmanex II (Hellma GmbH \& Co. KG),
whose volume concentration is 2 \%, at 35 $^{\circ}$C for 60 minutes. (2) dry
the tubes with a jet of gas using a canned spray (JBA-S481, Sanhayato Corp.).
(3) store them in an airtight container, until they were used for the
experiment. We avoided using ultrasonic bath because the cavitation process
seemed to create inhomogeneity on the inside surface of the tubes. Indeed,
when performing experiments with tubes cleaned with ultrasound, we noticed
irregular stick-and-slip motions. (In contrast, as shown in Fig. \ref{Fig1}
(c) below, all the invasion dynamics analyzed in this study were smooth, which
may indicate $\theta_{12}$ was very small.)

As shown in the series of snapshots reported in Fig. \ref{Fig1} (b), liquid 2
penetrates a tube immediately after the tube touches a lens of liquid 2 at
$t=0$, where liquid 2 gradually replaces liquid 1 - the visual contrast
between the liquids is amplified by the addition of a day (food red) in
ethanol. Pictures were taken with a digital camera (Nikon D800E with AFS-S
Micro Nikkor 60 mm 1.28 G ED) using the interval mode, and analyzed images
with \textit{Tracker} (The Open Source Physics Project). Typical results for
the invasion length $x$ as a function of time $t$ are given in Fig. \ref{Fig1} (c).

In most cases, the meniscus first propagates at a velocity of around 1 mm/s
that logically decreases when increasing the viscosity of liquid 1. In
addition, the meniscus accelerates significantly as it approaches the tube
end. This behavior, together with a case of a linear dynamics (represented by
green star symbols), markedly confronts with classical capillary impregnation,
for which the meniscus slows down with time.

\section{Model}

In the experiment, a tube is initially filled with liquid 1, at the cost of
the interfacial energy between liquid 1 and the tube wall, $\gamma_{1S}$ per
unit area. When the tube gets into contact with a lens of liquid 2, the liquid
starts to replace liquid 1 if the interfacial energy between liquid 2 and the
solid, $\gamma_{2S}$, is smaller than $\gamma_{1S}$, as in Fig. \ref{Fig1} (a)
and (b). This replacement suggests that liquid 1 does not form any thin film
during the invasion. Indeed, a film left by liquid 1 would cost the surface
energy $\gamma_{1S}+\gamma_{12\text{ }}$per unit area, an energy higher than
the initial state, denoting $\gamma_{12}$ as the interfacial energy between
liquids 1 and 2. The energy change per unit area as the contact line moves by
$dx$ is $dE=(\gamma_{2S}-\gamma_{1S})dx$, which gives a driving force
$f_{\gamma}=-dE/dx=\gamma_{1S}-\gamma_{2S}$ ($>0$). Hence, the force driving
the liquid invasion is given by%
\begin{equation}
F_{\gamma}=2\pi R\Delta\gamma\label{eqdr}%
\end{equation}
with%
\begin{equation}
\Delta\gamma=\gamma_{1S}-\gamma_{2S}%
\end{equation}
This can be written in the form $f_{\gamma}=\gamma_{12}\cos\theta_{12}$ on the
basis of a force balance at the contact line when the spreading coefficient
$\gamma_{1S}-\gamma_{2S}-\gamma_{12}$ is non positive. Here, $\theta_{12}$ is
the contact angle of a drop of liquid 2 on the solid when surrounded by liquid
1. In this case, the driving force is maximized for $\theta_{12}=0$, which
corresponds to a complete wetting condition for the invading liquid.

Motion is resisted by the Poiseuille viscous drag on tube walls, to which both
liquids 1 and 2 contribute:%
\begin{equation}
F_{\eta}=8\pi\lbrack\eta_{2}x+\eta_{1}(L-x)]\frac{dx}{dt} \label{eq6}%
\end{equation}
We assume that all other viscous and inertial effects can be neglected (see
Discussion for more details) and just consider the balance $F_{\gamma}%
=F_{\eta}$. As a result, we can completely solve the dynamics, which yields:%
\begin{equation}
\frac{\eta_{2}-\eta_{1}}{\eta_{1}}\frac{x}{L}=-1+\sqrt{1+(\eta_{2}-\eta
_{1})\frac{R\Delta\gamma t}{2\eta_{1}^{2}L^{2}}} \label{eq1}%
\end{equation}

Under the above assumption, this solution is valid irrespective of the
relative values of $\eta_{1}$ and $\eta_{2}$. Let us look at further
simplifications of this general expression. In the limit $\eta_{1}=\eta_{2}$,
this solution approaches the simple form:
\begin{equation}
x(t)=\frac{R\Delta\gamma}{4\eta_{1}L}t, \label{eq5}%
\end{equation}
The physics of this linear dynamics is clear: in this limit, the pre-existing
liquid is replaced by another invading liquid of same viscosity as invasion
proceeds, so that the viscous force remains constant during the invasion process.

More generally, the initial invading velocity $V$ is obtained from eq.
(\ref{eq1}). We get
\begin{equation}
V=\frac{R\Delta\gamma}{4\eta_{1}L} \label{eq4}%
\end{equation}
This expression for $V$ was recently derived in a previous study
\cite{walls2016capillary} as a limit of a non-explicit relation between $x$
and $t$ in a vertical geometry, where gravity matters.

By solving eq. (\ref{eq1}) at $x=L$ in terms of $t$, we also obtain a
theoretical prediction for the time $t_{L}$ for the complete replacement:%
\begin{equation}
t_{L}=\frac{2(\eta_{2}+\eta_{1})L^{2}}{R\Delta\gamma} \label{eq3}%
\end{equation}

Introduction of the renormalized length $X=\tilde{\eta}_{_{-}}x/L$ and time
$T=\tilde{\eta}_{_{+}}\tilde{\eta}_{_{-}}t/t_{L}$ with $\tilde{\eta}_{\pm
}=(\eta_{2}\pm\eta_{1})/\eta_{1}$ allows us to simplify eq. (\ref{eq1}) into a
universal form:%
\begin{equation}
X=-1+\sqrt{1+T} \label{eq2}%
\end{equation}

\section{Comparison between experiment and theory}

\begin{figure}[h]
\includegraphics[width=\textwidth]{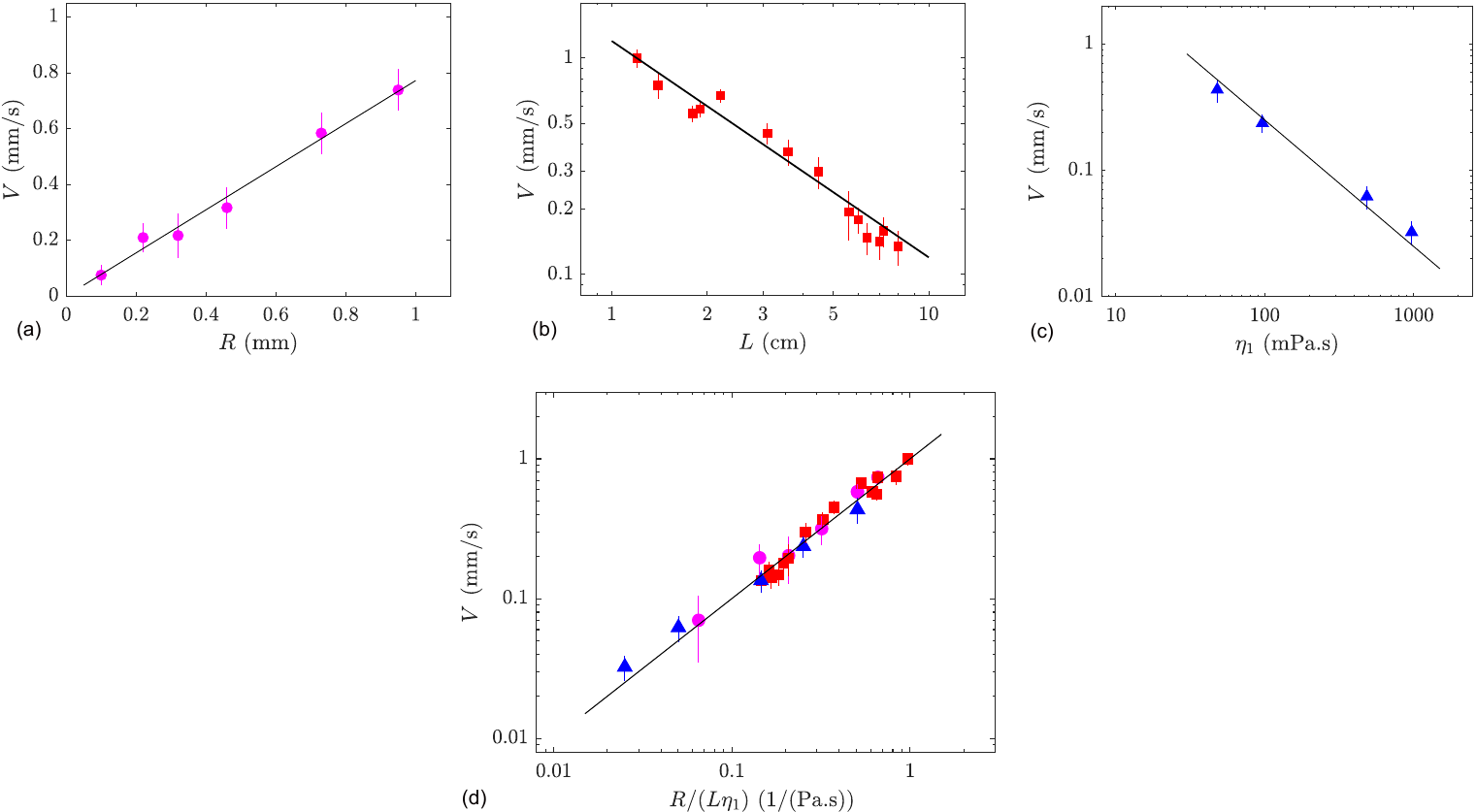}\caption{(a) Initial invasion
velocity $V$ as a function of $R$ with the other parameters fixed: $(\eta
_{1},\eta_{2},L)=(48,1.4,30)$. (b) $V$ as a function of $L$ with $(\eta
_{1},\eta_{2},R)=(48,1.4,0.56)$. (c) $V$ as a function of $\eta_{1}$ with
$(\eta_{2},L,R)=(1.4,30,0.73)$. $\eta_{1}$ and $\eta_{2}$ are given in
mPa$\cdot$s, and $L$ and $R$ are in mm. (d) $V$ as a function of the quantity
$R/(\eta_{1}L)$. Data are taken from (a) to (c), and all collapse in this
representation, as predicted by eq. (\ref{eq4}). The line represents the
relation $V=aR/(\eta_{1}L)$ with $a=1.1$ mN/m.}%
\label{Fig2}%
\end{figure}

These different predictions can be compared with the data. We first test eq.
(\ref{eq4}) by plotting the initial velocity $V$ of invasion as a function of
the tube radius $R$ (Fig. \ref{Fig2} a), of the tube length $L$ (Fig.
\ref{Fig2} b), and of the oil viscosity (Fig. \ref{Fig2} c). In all cases,
linear relationships are observed, as expected from eq. (\ref{eq4}). The
straight line that fits the data, goes through the origin in the linear plot
(a) and has a slope $-1$ in the log-log plots (b) and (c). All experiments can
be collected in a single plot as in Fig. \ref{Fig2}d to obtain an estimate of
$\Delta\gamma$. The data collapse on a line representing the relation
$V=aR/(\eta_{1}L)$ in agreement with eq. (\ref{eq4}), where $a$ is found to be
$a=1.1\pm0.1$ mN/m. This gives an estimate $\Delta\gamma=4.4\pm0.4$ mN/m
because eq. (\ref{eq4}) suggests $a=\Delta\gamma/4$.

\begin{figure}[h]
\includegraphics[width=\textwidth]{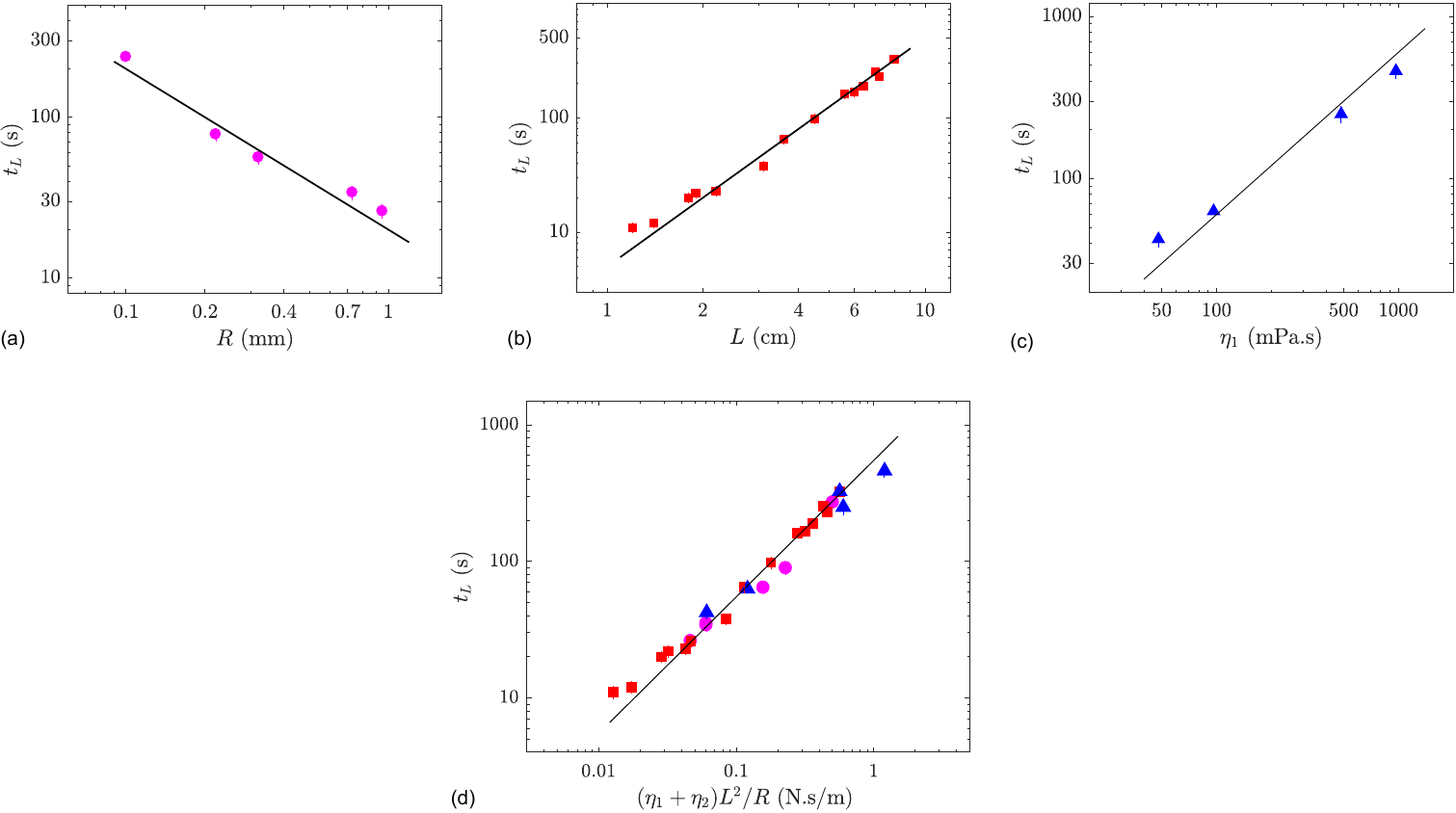}\caption{(a) Total time of
invasion $t_{L}$ as a function of $R$, keeping the other parameters fixed:
$(\eta_{1},\eta_{2},L)=(48,1.4,30)$. (b) $t_{L}$ as a function of $L$ with
$(\eta_{1},\eta_{2},R)=(48,1.4,0.56)$. (c) $t_{L}$ as a function of $\eta_{1}$
with $(\eta_{2},L,R)=(1.4,30,0.73).$ $\eta_{1}$and $\eta_{2}$ are given in
mPa$\cdot$s, and $L$ and $R$ are in mm. (d) $t_{L}$ as a function of the
quantity $(\eta_{1}+\eta_{2})L^{2}/R$. Data are taken from (a) to (c), and all
collapse in this representation, as predicted by eq. (\ref{eq3}). The line
represents the relation $t_{L}=b(\eta_{1}+\eta_{2})L^{2}/R$ with $1/b=1.8$
mN/m.}%
\label{Fig3}%
\end{figure}

We secondly test eq. (\ref{eq3}) in Fig. \ref{Fig3}. The dependences of
$t_{L}$ on $R$, $L$, and $\eta_{1}$ are confirmed in \ref{Fig3}a to c. All
data are collected in \ref{Fig3}d to obtain another estimate of $\Delta\gamma
$. The data collapse on a line representing $t_{L}=b(\eta_{1}+\eta_{2}%
)L^{2}/R$ in agreement with eq. (\ref{eq3}), where $1/b=1.8\pm0.3$ mN/m,
giving another estimate of $\Delta\gamma$: $\Delta\gamma=3.6\pm0.6$ mN/m.

The two estimates for $\Delta\gamma$ ($4.4\pm0.4$ and $3.6\pm0.6$ mN/m) are
consistent with each other, and are in agreement of our direct measurement of
$\Delta\gamma$, which was a few mN/m. The direct measurement showed
fluctuations. This is reasonable, because this quantity, i.e., $\gamma
_{1S}-\gamma_{2S}$, is sensitive to cleaning process. As a result, even though
a special care was taken, $\Delta\gamma$ fluctuates depending on the tube
used. We consider this is the main source for small but visible deviations of
the data from\ fitting lines in Figs. \ref{Fig2} and \ref{Fig3}. (The slight
deviation is small enough to demonstrate a good agreement between theory and
experiment, but the quality of the collapse onto a master curve is more
excellent in Fig. \ref{Fig4} below.) Note that quantitative determination of
the liquid-liquid surface tension such as $\gamma_{12}$ is still an issue of
current study \cite{berry2015measurement}.

However, this subtil issue of $\Delta\gamma$ is avoided to test the universal
dynamics predicted by eq. (\ref{eq2}), as explained below. The test result is
shown in Fig. \ref{Fig4}, with properly rescaling the distance $x$ and time
$t$. As expected, all data collapse, with significantly less deviations, on a
single master function $X(T)$ nicely fitted by eq. (\ref{eq2}).

\begin{figure}[h]
\includegraphics[width=0.4\textwidth]{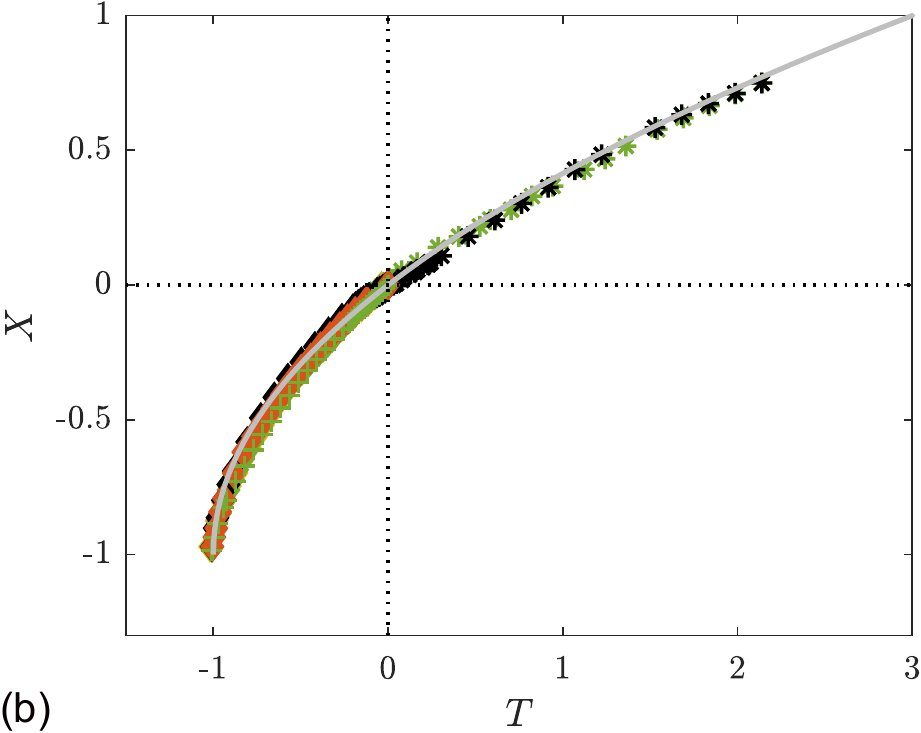}\caption{Data set from Fig.
\ref{Fig1} (c) replotted after rescaling of time and invasion length. All data
collapse, as expected from eq. (\ref{eq2}), which is shown by a line.}%
\label{Fig4}%
\end{figure}

As announced, in Fig. \ref{Fig4}, the analysis does not relay on the
measurements of $\Delta\gamma$, which is sensitive to cleaning process. All
the $\Delta\gamma$-dependences appear in eqs. (\ref{eq1}) and (\ref{eq2}) only
through $t_{L}$ (remind in particular the definitions $T=\tilde{\eta}_{_{+}%
}\tilde{\eta}_{_{-}}t/t_{L}$ and eq. (\ref{eq3})), while this quantity $t_{L}$
can be measured directly for each tube through the experiment. This means
that, if we use experimentally obtained $t_{L}$ to calculate the rescaled
variable $T$, we can escape a direct use of experimental values of
$\Delta\gamma$. Thus, the convincing collapse exhibited in Fig. \ref{Fig4}
indirectly justifies eq. (\ref{eq3}). As expected from above discussions, if
we instead use either of the two estimate of $\Delta\gamma$ to obtain Fig.
\ref{Fig4}, the quality of collapse of the data deteriorates distinctively.
Furthermore, $\Delta\gamma$ estimated from the oil of lowest viscosity (0.8
mPa$\cdot$s) was clearly different from the above two estimates. This could be
because of partial dissolution of the oil in ethanol. However, even in this
case, our analysis independent of $\Delta\gamma$ allowed a clear collapse onto
the master curve as demonstrated in Fig. \ref{Fig4}.

\section{Discussion}

In the raw data shown in Fig. \ref{Fig1}(c), most of the data sets show that
the dynamics accelerates as the invasion proceeds, as opposed to most of the
theories known for capillary invasion. In fact, in the contexts including the
imbibition of textured surfaces \cite{BicoEPL2001}, in addition to the
conventional Washburn dynamics in which $x\,$\ scales as $t^{1/2}$ in the
viscous regime \cite{IshinoOkumura2008}, the linear dynamics has been
discussed \cite{TaniPlosOne2014} and achieved \cite{delannoy2019dual}, and
other slowing dynamics in which $x$ scales as $t^{1/3}$ has been established
in a various different geometries
\cite{TangTang1994,JFM2011QuereCorners,ObaraPRER2012}.

However, the physical mechanism for this unusual behavior is very simple. When
the viscosity of the invading liquid $\eta_{2}$ is smaller than the previously
existing liquid $\eta_{1}$ ($\eta_{2}<\eta_{1}$), the total viscous drag is
reduced as the invasion proceeds as clear from the physics employed in eq.
(\ref{eq6}).

In the opposite case of $\eta_{2}>\eta_{1}$, we see the conventional
slowing-down dynamics. In fact, eq. (\ref{eq1}) is expressed as
\begin{equation}
\tilde{\eta}_{_{-}}\frac{x}{L}=\sqrt{\frac{\tilde{\eta}_{_{-}}RV_{0}t}{2L^{2}%
},}%
\end{equation}
which reduces to Washburn law in the limit $\eta_{2}\gg\eta_{1}$. The final
velocity $V_{f}=dx/dt$ at $t=t_{L}$ is given by%
\begin{equation}
V_{f}=\frac{\Delta\gamma Rt}{4\eta_{2}L} \label{eq7}%
\end{equation}
irrespective of relative magnitudes of $\eta_{1}$ and $\eta_{2}$. At short
times, the dynamics is always linear in $t$: $x=Vt$. In other words, eq.
(\ref{eq1}) predicts (i) the unusual accelerating dynamics when the invading
liquid is less viscous, (ii) the conventional slowing-down dynamics when the
invaded liquid is more viscous (as in the conventional case in which the
invaded fluid is air), and (iii) another unusual dynamics, which is linear in
time, when the two viscosities match with each other.

When discussing the universal equation in eq. (\ref{eq2}), it is worthwhile
noting that (i) when $\eta_{2}>\eta_{1}$, the terminal time for the invasion
$t=t_{L}$ corresponds to $T=1$, but (ii) when $\eta_{1}>\eta_{2}$, the
terminal time corresponds to $T=-1$. This means that in the master curve in
eq. (\ref{eq2}) shown in Fig. \ref{Fig2}(b), the dynamics proceeds from $T=0$
to $T=1$ for $\eta_{2}>\eta_{1}$, while it does from $T=0$ to $T=-1$ for
$\eta_{2}<\eta_{1}$. The variables $X$ and $T$ change their signs depending on
the relative magnitudes of $\eta_{1}$ and $\eta_{2}$.

The curvature effect of the lens of liquid 2 and inertia, which are neglected
in the present theory, can be justified. The thickness of the lens scaling as
$\sqrt{(-\gamma_{12}-\gamma_{2}+\gamma_{1})/(\Delta\rho g)}$ with $g$ the
gravitational acceleration is typically more than a few mm, which is
significantly larger than the tube radius, because of the small factor
$\Delta\rho=$ $\left\vert \rho_{2}-\rho_{1}\right\vert $. As for inertia,
Reynolds number Re for the present problem scales as $(\rho VR/\eta)(R/x)$
where $\rho$ and $\eta$ are the density and viscosity of the liquid in
question and $V$ is the characteristic velocity scale. The extra factor $R/x$
emerges in this case, because the inertia term in Navier-Stokes equation
scales as $\rho V^{2}/x$ since $V$ scales as $x/t$ in the present case. In the
present experiment, Re thus defined is always less than one for space and time
resolution of the present experiment, and thus it is logical that all the data
were successfully explained by the present theory.

We have neglected the following three viscous dissipations occurring at the
entrance and exit points of the tube and at the contact line, which will be
called the entrance, exit, and contact-line effects. The entrance effect is
associated with a viscous flow outside the tube, which develops in the volume
of the scale of tube radius around the entrance end of the tube, which scales
as $\eta V^{2}R$ with $\eta$ representing the larger of $\eta_{1}$ and
$\eta_{2}$. The exit effect is the counterpart of the entrance effect at the
exit, which scales as $\eta_{1}V^{2}R$. The contact-line effect is associated
with dissipation at the contact line, which scales as $\eta_{2}V^{2}R$.

To look into these three effects, experiments with longer capillary lengths
(thus we can use tubes with larger radius), which implies larger difference in
surface energy and smaller difference in density, would an interesting future
problem. In the present study, the agreement between theory and experiment
suggests that these effects can be neglected for the parameter range we
studied. However, we give further theoretical justification for $\eta_{1}%
>\eta_{2}$, because the discussion is rather simple in this case and this
condition is satisfied by most of the data in the present theory. The
Poiseuille dissipation considered in the present theory scales as $\eta
_{2}V^{2}x+\eta_{1}V^{2}(L-x)$. This is represented by $\eta_{1}V^{2}(L-x)$
for $\eta_{1}>\eta_{2}$. This implies the neglected three effects could matter
only at long times. Thus, the correction due to the three effects is
negligible for the initial dynamics represented by $V$ (Fig. \ref{Fig2}), for
the terminal time $t_{L}$ characterizing the overall dynamics (Fig.
\ref{Fig3}), and even for the entire dynamics represented by the $X-T$
relation with $X$ renormalized by $L$ (Fig. \ref{Fig4}).

The force balance leading to eq.(\ref{eq5}) and its integrated form were
already discussed in \cite{eley1946dynamical}. However, their interest is not
in the invasion dynamics itself but in developing a technique to exploit the
dynamics to determine the liquid-liquid surface tension. Accordingly, they did
not derive the explicit closed expression for the invasion length in
eq.(\ref{eq5}), and thus did not discuss the universal feature of the dynamics
elucidated in eq. (\ref{eq2}). This remarkable feature was systematically
confirmed by experiment in the present study. The discussion on the horizontal
geometry in \cite{eley1946dynamical} was very brief without showing any graphs
and the contamination was a great problem: the reproducible results were
obtained only when prewetting by the invading liquid was performed, which
corresponds to a special case of zero-contact angle. They were rather
interested in the capillary replacement in the vertical tube filled with
grains, but they could not obtain the data in a reproducible way, showing a
single graph obtained from "a good experiment," which was nonetheless rather
consistent with their theory for the tube with grains. Note that the issue of
contamination is resolved by the special cleaning process as already discussed.

\section{Dynamics with a slug\label{Sslug}}

\begin{figure}[h]
\includegraphics[width=\textwidth]{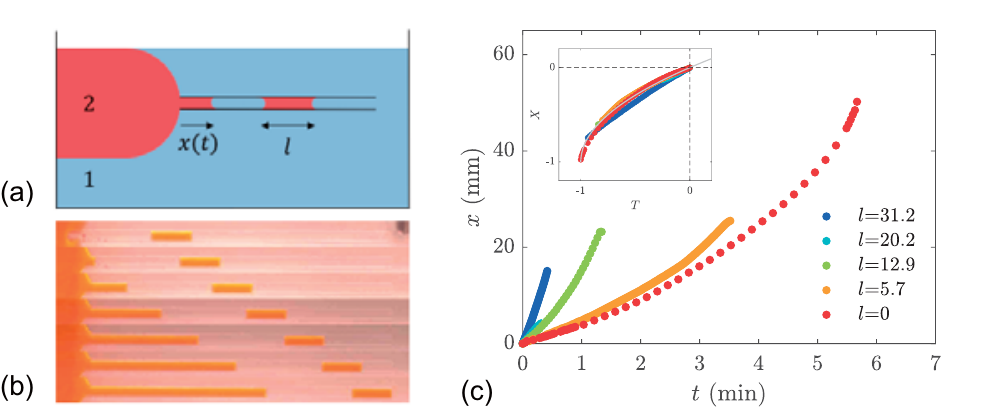}\caption{a) Schematic view of the
set up. Here, $l$ is the length of the slug/drop, $L$ the length of the tube,
$L^{\prime}$ the distance between the end of the tube and the right end of the
slug (i.e., the available distance for rise), and $x(t)$ the position of the
tip of invading bulk phase of ethanol, rising from $x=0$ to $x=L^{\prime}$. b)
Chronophotography of one experiment, with ethanol and silicon oil of viscosity
48 mPa$\cdot$s, in a glass tube of $(R,L)=(0.7,50)$ in mm. The interval
between two adjacent images is 68 s. c) The bulk invasion length $x$ vs $t$
for various slug length $l$ with $(R,L)=(0.7,100)$ in mm, obtained with using
the same liquid as in b. The inset shows the collapse of the data in the slug
experiment on the master curve $X-T$.}%
\label{Fig6}%
\end{figure}

In some practical situations, a small drop of the invading liquid could exist
before the bulk invasion starts, as in Fig. \ref{Fig6}(a). We consider the
case in which a slug of liquid 2 is already in the tube, and surrounded by
liquid 1. We call $l$ the slug's length, and $L^{\prime}$ its distance to the
free extremity of the tube, before the invasion starts (the slug initially
occupies the tube in the position from $x=L-(l+L^{\prime})$ to $x=L-L^{\prime
})$. In this case the invasion terminates when the front of the meniscus form
the bulk phase of liquid 2 approaches the position $x=L^{\prime}$, as shown in
Fig. \ref{Fig6}(a). This is because further displacement of the meniscus is
hindered by the pinning of the slug at the edge of the tube.

The dynamics in this case may still be described by the unified expression
given in eq. (\ref{eq2}), for modified definitions: $X=\tilde{\eta}_{_{-}%
}x/L_{s}$ and $T=\tilde{\eta}_{_{+}}\tilde{\eta}_{_{-}}t/t_{L,s}$. Here,
$L_{s}=L+\tilde{\eta}_{_{-}}l$ and $t_{L,s}$ is given by eq. (\ref{eq3}) with
$L$ replaced by $L_{s}$.\ The initial velocity given in eq. (\ref{eq4}) is
still valid if $L$ is replaced by $L_{s}$. The universal expression, eq.
(\ref{eq2}), can also be expressed by the invasion time $t_{L}^{\prime}$ in
this case, i.e., the time it takes for the new meniscus to go from $x=0$ to
$x=L^{\prime}$. This is because $t_{L,s}$ satisfies the relation
$t_{L}^{\prime}=t_{L,s}((1+L^{\prime}/(l+L/\tilde{\eta}_{_{-}}))^{2}-1)$. The
systematic comparison of this theoretical prediction with experiment is shown
in Fig. \ref{Fig6}(c), which shows a reasonable data collapse onto the master curve.

\section{Conclusion}

In the present study, we focused on the capillary replacement in the
horizontal geometry, when the viscosity of the pre-existing liquid is
dominant. We solved the dynamics giving explicitly the functional dependence
of the invaded length $x$ on the elapsed time $t$. The explicit solution
elucidates a universal feature of the replacement dynamics: the analytical
expression unifies three distinctive dynamics, including non-conventional
linear and accelerating dynamics, in addition to the conventional slowing
dynamics. Furthermore, we derived a simple analytical result for the complete
replacement time, the time required for the second liquid to completely
replace the pre-existing liquid, which has never been discussed. We performed
systematic experiments to test our theory and showed clear agreements between
the theory and our data on the universal dynamics and the complete replacement
time. We also experimentally confirm the initial invading velocity, which can
be derived from our theory. Although the theory for the initial velocity has
been available in the literature, a systematic confirmation has been lacking
previously. The issues discussed in the present study should be significantly
important for our fundamental understanding of wetting phenomena, and may be
relevant to applications in various fields, which include microfluidics
\cite{SquiresQuake2005} and petroleum industry \cite{HeleShawPetroleum2010}.

\begin{acknowledgments}
%K. O. is grateful to Mitsubishi Chemical Corporation for having discussions that happened to drive him to start this study. 
J.A is grateful to David
Qu\'{e}r\'{e} for hosting her in his lab, in collaboration with K.O, to work
on an interesting subject, which gave birth to Sec. \ref{Sslug} of the present
article. K.O and J.A also thank Qu\'{e}r\'{e} for his remarks, experimental as
well as theoretical, and for critical reading of drafts of the present
article. This work was partly supported by JSPS\ KAKENHI Grant Number
JP19H01859. J. A. thanks SAINT GOBAIN for financial support ("Innovative
Process \& Materials," l'X - Ecole polytechnique and the Fondation de l'Ecole polytechnique).
\end{acknowledgments}

%\bibliography{fracture}
%%\bibliographystyle{naturemagMy}
%%\bibliography{C:/Users/okumura/Documents/main/JabRef/granular,C:/Users/okumura/Documents/main/JabRef/fracture,C:/Users/okumura/Documents/main/JabRef/wetting}

\begin{thebibliography}{99}                                                                                               %
\expandafter\ifx\csname url\endcsname\relax


\fi
\expandafter\ifx\csname urlprefix\endcsname\relax

\fi
\providecommand{\bibinfo}[2]{#2} \providecommand{\eprint}[2][]{\url{#2}}

\bibitem {de2013capillarity}\bibinfo{author}{De~Gennes, P.-G.},
\bibinfo{author}{Brochard-Wyart, F.} \&  \bibinfo{author}{Qu{\'e}r{\'e}, D.}
\newblock \emph{\bibinfo{title}{Capillarity and wetting phenomena: drops,
bubbles, pearls, waves}} (\bibinfo{publisher}{Springer Science \& Business
Media}, \bibinfo{year}{2013}).

\bibitem {hultmark2011influence}\bibinfo{author}{Hultmark, M.},
\bibinfo{author}{Aristoff, J.~M.} \&  \bibinfo{author}{Stone, H.~A.}
\newblock \bibinfo{title}{The influence of the gas phase on liquid imbibition
in capillary tubes}.
\newblock \emph{\bibinfo{journal}{Journal of Fluid Mechanics}}
\textbf{\bibinfo{volume}{678}}, \bibinfo{pages}{600--606}  (\bibinfo{year}{2011}).

\bibitem {elliott1967dynamic}\bibinfo{author}{Elliott, G.} \&
\bibinfo{author}{Riddiford, A.}
\newblock \bibinfo{title}{Dynamic contact angles: I. the effect of impressed
motion}.
\newblock \emph{\bibinfo{journal}{Journal of colloid and interface science}}
\textbf{\bibinfo{volume}{23}}, \bibinfo{pages}{389--398}  (\bibinfo{year}{1967}).

\bibitem {hansen1971dynamic}\bibinfo{author}{Hansen, R.~J.} \&
\bibinfo{author}{Toong, T.}
\newblock \bibinfo{title}{Dynamic contact angle and its relationship to forces
of hydrodynamic origin}.
\newblock \emph{\bibinfo{journal}{Journal of Colloid and Interface Science}}
\textbf{\bibinfo{volume}{37}}, \bibinfo{pages}{196--207}  (\bibinfo{year}{1971}).

\bibitem {barenblatt1990Rocks}\bibinfo{author}{Barenblatt, G.},
\bibinfo{author}{Entov, V.} \&  \bibinfo{author}{Ryzhik, V.}
\newblock \emph{\bibinfo{title}{Theory of Fluid Flows Through Natural Rocks}}
(\bibinfo{publisher}{Spreinger}, \bibinfo{year}{1990}).

\bibitem {gerritsen2005modeling}\bibinfo{author}{Gerritsen, M.~G.} \&
\bibinfo{author}{Durlofsky, L.~J.}
\newblock \bibinfo{title}{Modeling fluid flow in oil reservoirs}.
\newblock \emph{\bibinfo{journal}{Annu. Rev. Fluid Mech.}}
\textbf{\bibinfo{volume}{37}}, \bibinfo{pages}{211--238}  (\bibinfo{year}{2005}).

\bibitem {eley1946dynamical}\bibinfo{author}{Eley, D.} \&
\bibinfo{author}{Pepper, D.}
\newblock \bibinfo{title}{A dynamical determination of adhesion tension}.
\newblock \emph{\bibinfo{journal}{Transactions of the Faraday Society}}
\textbf{\bibinfo{volume}{42}}, \bibinfo{pages}{697--702}  (\bibinfo{year}{1946}).

\bibitem {mumley1986kinetics}\bibinfo{author}{Mumley, T.~E.},
\bibinfo{author}{Radke, C.} \&  \bibinfo{author}{Williams, M.~C.}
\newblock \bibinfo{title}{Kinetics of liquid/liquid capillary rise: I.
experimental observations}.
\newblock \emph{\bibinfo{journal}{Journal of colloid and interface science}}
\textbf{\bibinfo{volume}{109}}, \bibinfo{pages}{398--412}  (\bibinfo{year}{1986}).

\bibitem {walls2016capillary}\bibinfo{author}{Walls, P.~L.},
\bibinfo{author}{Dequidt, G.} \&  \bibinfo{author}{Bird, J.~C.}
\newblock \bibinfo{title}{Capillary displacement of viscous liquids}.
\newblock \emph{\bibinfo{journal}{Langmuir}} \textbf{\bibinfo{volume}{32}},
\bibinfo{pages}{3186--3190} (\bibinfo{year}{2016}).

\bibitem {budaraju2016capillary}\bibinfo{author}{Budaraju, A.},
\bibinfo{author}{Phirani, J.},  \bibinfo{author}{Kondaraju, S.} \&
\bibinfo{author}{Bahga, S.~S.}
\newblock \bibinfo{title}{Capillary displacement of viscous liquids in
geometries with axial variations}.
\newblock \emph{\bibinfo{journal}{Langmuir}} \textbf{\bibinfo{volume}{32}},
\bibinfo{pages}{10513--10521} (\bibinfo{year}{2016}).

\bibitem {berry2015measurement}\bibinfo{author}{Berry, J.~D.},
\bibinfo{author}{Neeson, M.~J.},  \bibinfo{author}{Dagastine, R.~R.},
\bibinfo{author}{Chan, D.~Y.} \&  \bibinfo{author}{Tabor, R.~F.}
\newblock \bibinfo{title}{Measurement of surface and interfacial tension using
pendant drop tensiometry}.
\newblock \emph{\bibinfo{journal}{Journal of colloid and interface science}}
\textbf{\bibinfo{volume}{454}}, \bibinfo{pages}{226--237}  (\bibinfo{year}{2015}).

\bibitem {BicoEPL2001}\bibinfo{author}{Bico, J.},
\bibinfo{author}{Tordeux, C.} \&  \bibinfo{author}{Qu\'{e}r\'{e}, D.}
\newblock \bibinfo{title}{Rough wetting}.
\newblock \emph{\bibinfo{journal}{Europhys. Lett.}}
\textbf{\bibinfo{volume}{55}}, \bibinfo{pages}{214--220}  (\bibinfo{year}{2001}).

\bibitem {IshinoOkumura2008}\bibinfo{author}{Ishino, C.} \&
\bibinfo{author}{Okumura, K.}
\newblock \bibinfo{title}{Wetting transitions on textured hydrophilic
surfaces}. \newblock \emph{\bibinfo{journal}{Eur. Phys. J. E}}
\textbf{\bibinfo{volume}{25}}, \bibinfo{pages}{415--424}  (\bibinfo{year}{2008}).

\bibitem {TaniPlosOne2014}\bibinfo{author}{Tani, M.} \emph{et~al.}
\newblock \bibinfo{title}{Capillary rise on legs of a small animal and on
artificially textured surfaces mimicking them}.
\newblock \emph{\bibinfo{journal}{Plos One}} \textbf{\bibinfo{volume}{9}},
\bibinfo{pages}{e96813} (\bibinfo{year}{2014}).

\bibitem {delannoy2019dual}\bibinfo{author}{Delannoy, J.},
\bibinfo{author}{Lafon, S.},  \bibinfo{author}{Koga, Y.},
\bibinfo{author}{Reyssat, {\'E}.} \&  \bibinfo{author}{Qu{\'e}r{\'e}, D.}
\newblock \bibinfo{title}{The dual role of viscosity in capillary rise}.
\newblock \emph{\bibinfo{journal}{Soft matter}} \textbf{\bibinfo{volume}{15}}%
,  \bibinfo{pages}{2757--2761} (\bibinfo{year}{2019}).

\bibitem {TangTang1994}\bibinfo{author}{{Lei-Han Tang}} \&
\bibinfo{author}{{Yu Tang}}.
\newblock \bibinfo{title}{Capillary rise in tubes with sharp grooves}.
\newblock \emph{\bibinfo{journal}{J. Phys. II France}}
\textbf{\bibinfo{volume}{4}}, \bibinfo{pages}{881--890}  (\bibinfo{year}{1994}).

\bibitem {JFM2011QuereCorners}\bibinfo{author}{Ponomarenko, A.},
\bibinfo{author}{Qu\'{e}r\'{e}, D.} \&  \bibinfo{author}{Clanet, C.}
\newblock \bibinfo{title}{A universal law for capillary rise in corners}.
\newblock \emph{\bibinfo{journal}{J. Fluid Mech.}}
\textbf{\bibinfo{volume}{666}}, \bibinfo{pages}{146--154}  (\bibinfo{year}{2011}).

\bibitem {ObaraPRER2012}\bibinfo{author}{Obara, N.} \&
\bibinfo{author}{Okumura, K.}
\newblock \bibinfo{title}{Imbibition of a textured surface decorated by short
pillars with rounded edges}. \newblock \emph{\bibinfo{journal}{Phys. Rev. E}}
\textbf{\bibinfo{volume}{86}},  \bibinfo{pages}{020601(R)} (\bibinfo{year}{2012}).

\bibitem {SquiresQuake2005}\bibinfo{author}{Squires, T.~M.} \&
\bibinfo{author}{Quake, S.~R.}
\newblock \bibinfo{title}{Microfluidics: Fluid physics at the nanoliter scale}.
\newblock \emph{\bibinfo{journal}{Rev. Mod. Phys.}}
\textbf{\bibinfo{volume}{77}}, \bibinfo{pages}{977} (\bibinfo{year}{2005}).

\bibitem {HeleShawPetroleum2010}\bibinfo{author}{Shad, S.},
\bibinfo{author}{Salarieh, M.},  \bibinfo{author}{Maini, B.} \&
\bibinfo{author}{Gates, I.~D.}
\newblock \bibinfo{title}{The velocity and shape of convected elongated liquid
drops in narrow gaps}.
\newblock \emph{\bibinfo{journal}{J. Petroleum Sci. Eng.}}
\textbf{\bibinfo{volume}{72}}, \bibinfo{pages}{67--77}  (\bibinfo{year}{2010}).
\end{thebibliography}

\end{document}